\documentclass[journal,twoside,web]{ieeecolor}
\usepackage{generic}
\usepackage{cite}
\usepackage{amsmath,amssymb,amsfonts}
\usepackage{algorithmic}
\usepackage{graphicx}
\usepackage{textcomp}
\usepackage[utf8]{inputenc}
\usepackage{textcomp}
\usepackage{subfig}
\usepackage[version=4]{mhchem}
\usepackage{siunitx}
\usepackage{longtable,tabularx}
\usepackage{color}
\usepackage{cancel}
\usepackage{gensymb}
\usepackage{mathtools}
\usepackage{bm}
\usepackage{amsthm}
\newtheorem{Dumm}{Dummy}[section]
\newtheorem{Definition}[Dumm]{Definition}
\newtheorem{Property}[Dumm]{Property}
\DeclareMathOperator{\diag}{diag} 
\newcommand{\mybar}[1]{\mkern 1.5mu\overline{\mkern-1.5mu#1\mkern-1.5mu}\mkern 1.5mu}

\def\BibTeX{{\rm B\kern-.05em{\sc i\kern-.025em b}\kern-.08em
    T\kern-.1667em\lower.7ex\hbox{E}\kern-.125emX}}
\begin{document}
\title{Linear Fractional Transformation modeling of multibody dynamics around parameter-dependent equilibrium}
\author{Ervan Kassarian, Francesco Sanfedino, Daniel Alazard, Charles-Antoine Chevrier, Johan Montel
\thanks{Submitted for review on 19/07/2021. This work was funded by ISAE-SUPAERO and CNES (French space agency -- grant 51/18660).}
\thanks{E. Kassarian, F. Sanfedino, and D. Alazard are with ISAE-Supaero, Toulouse, 31055 France (e-mails: ervan.kassarian@isae.fr, francesco.sanfedino@isae.fr, daniel.alazard@isae.fr).}
\thanks{J. Montel and C.A. Chevrier are with CNES, Toulouse, 31055 France (e-mails: charlesantoine.chevrier@cnes.fr, johan.montel@cnes.fr).}
}

\maketitle

\begin{abstract}
This paper proposes a new Linear Fractional Transformation (LFT) modeling approach for uncertain Linear Parameter Varying (LPV) multibody systems with parameter-dependent equilibrium.
Traditional multibody approaches, which consist in building the nonlinear model of the whole structure and linearizing it around equilibrium after a numerical trimming, do not allow to isolate parametric variations with the LFT form. Although additional techniques, such as polynomial fitting or symbolic linearization, can provide an LFT model, they may be time-consuming or miss worst-case configurations.
The proposed approach relies on the trimming and linearization of the equations at the substructure level, before assembly of the multibody structure, which allows to only perform operations that preserve the LFT form throughout the linearization process. 
Since the physical origin of the parameters is retained, the linearized LFT-LPV model of the structure exactly covers all plants, in a single parametric model, without introducing conservatism or fitting errors.
An application to the LFT-LPV modeling of a robotic arm is proposed; in its nominal configuration, the model obtained with the proposed approach matches the model provided by the software \textit{Simscape Multibody}, but it is enhanced with parametric variations with the LFT form; a robust LPV synthesis is performed using Matlab \textit{robust control toolbox} to illustrate the capacity of the proposed approach for control design.
\end{abstract}



\section*{Nomenclature}

\begin{table}[hbt!]
\centering
\begin{tabular}{ll}
$(^*\mathbf u)$ & Skew symmetric matrix of vector $\mathbf u$, such that $\mathbf u \times \mathbf{v}=(^*\mathbf u)\mathbf{v}$ \\
$\left[\mathbf{X}\right]_{R_x}$ & $\mathbf{X}$ (vector or tensor) projected in frame $R_x$\\
$\mybar{\mathbf X}$ & $\mathbf x$ (scalar or vector) evaluated at equilibrium\\
$\delta \mathbf X$ & First-order variations of $\mathbf x$ around equilibrium\\
\end{tabular}
\end{table}

\section{Introduction}

Multibody systems have applications in various fields such as aeronautics, aerospace or robotics, with numerous modeling and formulation approaches \cite{Rong2019a}.
Even though multibody dynamics are inherently nonlinear, it is often useful to linearize them at equilibrium to study the stability, perform modal analysis or apply classical linear control methods.
In practical engineering problems, many parameter uncertainties impact the dynamics of the system and must be taken into account for robust analysis and control.
When working on the uncertain linear model, a representation of the uncertainties with a bounded and unknown operator $\Delta$, based on the Linear Fractional Transformation (LFT), enables powerful tools to perform worst-case robust analysis and control such as $\mu$-analysis or $\mathcal H_{\infty}$-synthesis \cite{Zhou1996}.
Furthermore, the nonlinear system can often be approximated by a Linear Parameter Varying (LPV) model around a slowly-varying equilibrium, where the varying or nonlinear terms are also represented in the operator $\Delta$ of the LFT. Finally, some mechanical parameters, such as masses of some elements, may be considered as decision variables and included in the LFT to be optimized simultaneously with the controller in multidisciplinary co-design approaches. In this paper, the uncertain, varying and decision parameters are referred to as parameters of interest.
Classical multibody approaches, consisting in building the nonlinear model of the structure by assembly of the individual models and then linearizing this nonlinear model around equilibrium, are unable to directly provide the LFT-LPV model. This is due to the trim conditions depending on the parameters of interest: for example, consider the small angles variations of a pendulum with an uncertain mass -- the gravity introduces a stiffness which depends on the uncertain mass, and a numerical trimming preceding the linearization will only capture a single parametric configuration of this stiffness rather than a parameterized LFT model taking into account the parametric uncertainty on the mass.
For a more general class of systems, the use of symbolic linearization was proposed in \cite{Marcos2005,Szabo2011} to overcome this issue, but it is computationally costly for complex systems, especially when dealing with many parameters or high-order dynamics. Consequently, the most common practice for systems with parameter-dependent trim conditions is to perform numerical linearizations around a grid of equilibrium points corresponding to particular values of the parameters, and to generate a model covering all Linear Time Invariant (LTI) models of the grid using multivariable polynomial fitting techniques \cite{Pfifer2011,Roos2014}. However, this procedure may introduce conservatism or miss worst-case configurations, and may be time consuming when there are many parameters or when a fine grid is required.

For the modeling of large space structures such as satellites with flexible solar panels in micro-gravity conditions, a general framework was introduced in \cite{alazard2008linear}, and implemented in a generic toolbox named \emph{Satellite Dynamics Toolbox} (SDT) \cite{Alazard2020}, to build linear models of flexible multibody systems. Based on Newton-Euler equations, this tool allows to build the dynamic model of the whole structure by assembling the individual models of each substructure based on the Two-Input Two-Output (TITOP) formalism \cite{Alazard2015}. Some assets of this approach include the compliance with various substructure models and boundary conditions, and support of the interfacing with finite element software when the model includes complex substructures \cite{Sanfedino2018}. The resulting model is provided under the form of a block-diagram with minimal number of states, and the parameters can be isolated to obtain a minimal LFT model, allowing robust control \cite{Sanfedino2021} or integrated control/structure co-design with the $\mathcal H_{\infty}$ synthesis \cite{Perez2016}.

In this paper, the framework from \cite{alazard2008linear,Alazard2020,Alazard2015,Sanfedino2018,Perez2016,Sanfedino2021}
is extended to the modeling of multibody systems undergoing variations around a uniformly accelerated motion, e.g. for systems subject to gravity (robotic arms, aircrafts, civil machinery, stratospheric balloons...) or space systems during a thrust phase (launchers, spacecrafts).
It was motivated by the need for robust control for stratospheric balloons, which are complex multibody systems subject to gravity with uncertain masses \cite{Kassarian2021} that cannot be modeled with current multibody software due to the parameter-dependent trim conditions.
Rather than linearizing the nonlinear model of the multibody structure, the proposed approach linearizes each individual substructure and kinematic joint and assembles them to build the LFT model of the structure, after an analytical computation of the parameter-dependent equilibrium.
It allows to only perform operations that preserve the LFT form throughout the linearization process.
The LFT model regroups all parametric configurations in one single model, enabling modern analysis and control tools like $\mu$-analysis or $\mathcal H_{\infty}$-synthesis, and is obtained without resorting to symbolic trimming of the nonlinear model or polynomial fitting of a set of LTI plants; in particular, the LFT model exactly covers all plants within the specified bounds without introducing conservatism or fitting error. Since the linearization procedure only relies on basic block-diagram manipulations, the LFT model is obtained in a reasonable amount of time.
From the control engineer's point of view, the proposed approach can be implemented in Matlab-Simulink to build complex multibody structures by interconnecting the individual bodies.
Targeted engineering applications include modeling of uncertain LPV multibody systems and lumped-parameter modeling of uncertain flexible systems, for the purpose of robust control, gain scheduling, vibrations control, or integrated control/structure co-design, along with control design tools such as Matlab \textit{robust control toolbox}.
To the authors knowledge, this approach is the first contribution addressing the parametric model linearization around parameter-dependent equilibrium in the general context of uncertain multibody systems.

The dynamics of rigid bodies are modeled with Newton-Euler equations in Section \ref{sec:2_rigid_body}, and the equations of the revolute joint are presented in Section \ref{sec:3_interconnection}. Section \ref{sec:4_linearization} discusses the equilibrium and the linearization of the individual models of rigid body and revolute joint and the compatibility with LFT formalism. The assembly, trim and linearization algorithm, allowing to keep the LFT dependency of the model on the parameters of interest during the linearization, is detailed in Section \ref{sec:5_procedure}.
Finally, Section \ref{sec:6_application} presents an application to the LFT modeling of an LPV robotic arm; the model is validated with a comparison to \textit{Simscape Multibody} and a LPV control design is performed to illustrate the capacity of the proposed approach for control design.

\section{Rigid body dynamics}

\label{sec:2_rigid_body}

\subsection{Description of the motion}

\label{sec:2A}

\begin{Definition}\label{def:acceleration} Uniformly accelerated reference frame $\mathcal R$ \newline
Let $\mathcal R = (O,\mathbf{x},\mathbf{y},\mathbf{z})$ be a reference frame in uniform acceleration, represented by the $3\times1$ vector $\mathbf a$, with regard to an inertial reference frame $\mathcal R_i$. 
\end{Definition}

In this paper, the motion is described in the reference frame $\mathcal R$.
This equilibrium condition can represent a gravity field or an acceleration during a thrust phase for a space system. 

\begin{Definition}\label{def:def1} Motion in the reference frame $\mathcal R$ \newline
     Let us define the following vectors:
    \begin{itemize}
        \item $\boldsymbol{\mathbf x}^{\mathcal B}_P= \begin{bsmallmatrix} OP \\ \boldsymbol{\theta}^\mathcal{B}	\end{bsmallmatrix}$ the $6\times1$ pose vector of body $\mathcal B$ at point $P$, with $OP$ the $3\times1$ position vector of $P$ and $\boldsymbol{\theta}^\mathcal{B}$ the $3\times1$ vector of Euler angles of $\mathcal B$ with regard to $\mathcal R$.
        \item $\boldsymbol{\mathbf x'}^{\mathcal B}_P= \begin{bsmallmatrix} \mathbf{v}^{\mathcal B}_P\\ \boldsymbol{\omega}^{\mathcal B} \end{bsmallmatrix}$ the $6\times1$ dual velocity vector of body $\mathcal B$ at point $P$, with $\mathbf{v}^{\mathcal B}_P = \left.\frac{d \overrightarrow{OP}}{dt}\right|_{\mathcal R}$ and $\boldsymbol{\omega}^{\mathcal B}$ the angular velocity of $\mathcal B$ with regard to $\mathcal R$.
        \item $\boldsymbol{\mathbf x''}^{\mathcal B}_P=\left.\frac{d \boldsymbol{\mathbf x'}^{\mathcal B}_P}{dt}\right|_{\mathcal R}= \begin{bsmallmatrix} \dot{\mathbf{v}}^{\mathcal B}_P \\ \boldsymbol{\dot{\omega}}^{\mathcal B}\end{bsmallmatrix}$ the $6\times1$ dual acceleration vector of body $\mathcal B$ at point $P$.
        \item $\mathbf{m}^{\mathcal B}_P = [ \mathbf a^T \, , \, (\boldsymbol{\mathbf x''}^{\mathcal B}_P)^T \, , \, (\boldsymbol{\mathbf x'}^{\mathcal B}_P)^T \, , \, (\boldsymbol{\mathbf x}^{\mathcal B}_P)^T ]^T$ is defined as the motion vector of body $\mathcal B$ at point $P$.
    \end{itemize}
\end{Definition}

Noting $\mathbf a_{\mathrm 6}= \begin{bsmallmatrix} \boldsymbol{\mathbf a}\\ \mathbf 0_{3\times1} \end{bsmallmatrix}$, the linear and angular accelerations of body $\mathcal B$ at point $P$ with respect to $\mathcal R_i$ are:
\begin{equation} \label{eq:inertial_acceleration}
\left[\begin{array}{c} \left.\mathbf{a}^\mathcal{B}_P \right|_{\mathcal R_i}\\  \left.\boldsymbol{\dot{\omega}}^\mathcal{B}\right|_{\mathcal R_i}\end{array}\right] = \boldsymbol{\mathbf x''}^{\mathcal B}_P + \mathbf a_{\mathrm 6} \;.
\end{equation}

\subsection{Newton-Euler equations for rigid bodies}

Let us consider a body $\mathcal B$ of mass $m^{\mathcal{B}}$ and matrix of inertia $\mathbf{J}^{\mathcal{B}}_B$ at center of gravity $B$.
Newton-Euler equations read at $B$:
\begin{equation} \label{eq:Newton_euler_b}
\resizebox{\hsize}{!}{$
   \underbrace{\left[\begin{array}{c}\mathbf{F}^{\mathcal B} \\ \mathbf{T}^{\mathcal B}_{B}\end{array}\right]}_{\mathbf W^{\mathcal B}_{B}} = \underbrace{\left[\begin{array}{cc}m^{\mathcal{B}}\mathbf I_3 &\mathbf{0}_{3\times 3}\\\mathbf{0}_{3\times 3} & \mathbf{J}^{\mathcal{B}}_B \end{array}\right]}_{\mathbf{D}^{\mathcal{B}}_B} 
   \left[\begin{array}{c} \left.\mathbf{a}^\mathcal{B}_B \right|_{\mathcal R_i}\\  \left.\boldsymbol{\dot{\omega}}^\mathcal{B}\right|_{\mathcal R_i}\end{array}\right]
   + \left[\begin{array}{c}\mathbf{0}_{3\times 1}\\ (^*\boldsymbol\omega^{\mathcal B})\mathbf{J}^{\mathcal{B}}_B\boldsymbol\omega^{\mathcal B} \end{array}\right] $}
\end{equation}
where $\mathbf W^{\mathcal B}_{B} = \begin{bsmallmatrix}  \mathbf{F}^{\mathcal B} \\ \mathbf{T}^{\mathcal B}_{B} \end{bsmallmatrix}$ is the $6\times1$ wrench vector (force $\mathbf{F}^{\mathcal B}$ and torque $\mathbf{T}^{\mathcal B}_{B}$) applied to the body $\mathcal B$ at point $B$. Definition \ref{def:tau_BP} and property \ref{prop:kinematic_transport} were introduced in \cite{alazard2008linear} to transport equation \eqref{eq:Newton_euler_b} to any other point $P$ of the body $\mathcal B$.

\begin{Definition}\label{def:tau_BP} Kinematic model \cite{alazard2008linear} \newline
	The $6\times6$ tensor
	$\boldsymbol\tau_{PC} = \begin{bsmallmatrix} \mathbf I_3 & (^*\overrightarrow{PC})\\\mathbf{0}_{3\times 3} & \mathbf I_3 \end{bsmallmatrix} $
	is defined as the kinematic model between two points $P$ and $C$.
\end{Definition}

\begin{Property}{Transport of the vectors \cite{alazard2008linear}: }\label{prop:kinematic_transport}
\begin{itemize}
	\item Dual velocity vector: $\boldsymbol{\mathbf x'}^{\mathcal B}_P = \boldsymbol\tau_{PC} \boldsymbol{\mathbf x'}^{\mathcal B}_C$
	\item Dual acceleration vector: $$\boldsymbol{\mathbf x''}^{\mathcal B}_P = \boldsymbol\tau_{PC} \boldsymbol{\mathbf x''}^{\mathcal B}_C + \begin{bsmallmatrix} (^*\bm \omega^{\mathcal B}) (^*\overrightarrow{PC})\bm \omega^{\mathcal B}\\ \mathbf{0}_{3\times 1}\end{bsmallmatrix}$$
	\item Wrench vector: $\mathbf W^{\mathcal B}_{C} = \boldsymbol\tau^T_{PC} \mathbf W^{\mathcal B}_{P}$ 
	\item Inverse kinematic model: $\boldsymbol\tau_{PC}^{-1}=\boldsymbol\tau_{CP}$ 
	\item Transitivity: $\boldsymbol\tau_{PC}\boldsymbol\tau_{CP'}=\boldsymbol\tau_{PP'}$.
\end{itemize}	
\end{Property}

Using property \ref{prop:kinematic_transport} to transport the vectors from point $B$ to another point $P$ of body $\mathcal B$, and since $\boldsymbol\tau_{BP} \mathbf a_{\mathrm 6} = \mathbf a_{\mathrm 6}$, equation \eqref{eq:Newton_euler_b} is transported to $P$:
\begin{equation} \label{eq:Newton_euler}
\resizebox{\hsize}{!}{$
   \mathbf W^{\mathcal B}_P = \underbrace{\boldsymbol\tau_{BP}^T \mathbf{D}^{\mathcal{B}}_B\boldsymbol\tau_{BP}}_{\mathbf{D}^{\mathcal{B}}_P}
   \left(  \boldsymbol{\mathbf x''}^{\mathcal B}_P + \mathbf a_{\mathrm 6} \right)
   + \underbrace{\begin{bsmallmatrix} m^{\mathcal{B}} (^*\bm \omega^{\mathcal B})(^*\overrightarrow{BP})\boldsymbol\omega^{\mathcal B}  \\ (^*\boldsymbol\omega^{\mathcal B})\left(\mathbf{J}^{\mathcal{B}}_B -m^{\mathcal{B}}\,(^*\overrightarrow{BP})^2 \right)\boldsymbol\omega^{\mathcal B}\end{bsmallmatrix}}_{\mathrm{NL}(P, \boldsymbol\omega^{\mathcal B})} $}
\end{equation}
where $\mathrm{NL}(P, \boldsymbol\omega^{\mathcal B})$ regroups the nonlinear terms, and $\mathbf{D}^{\mathcal{B}}_P$ is defined as the direct dynamics model of body $\mathcal B$ at point $P$.

\subsection{Projection in the body's frame}

In order to describe each body independently from the others, equation \eqref{eq:Newton_euler} is projected in the reference frame $\mathcal R_b$ attached to $\mathcal B$:
\begin{equation} \label{eq:NE_short_g_projected}
    [\mathbf W^{\mathcal B}_P]_{\mathcal R_b} = [\mathbf{D}^{\mathcal{B}}_P]_{\mathcal R_b} \left( [\boldsymbol{\mathbf x''}^{\mathcal B}_P]_{\mathcal R_b} + [\mathbf a_{\mathrm 6}]_{\mathcal R_b} \right) + [\mathrm{NL}(P, \boldsymbol\omega^{\mathcal B})]_{\mathcal R_b} \; .
\end{equation}
The kinematic and direct dynamic models are conveniently written in the body's frame. The inertial uniform acceleration vector $\mathbf a$ is defined in the inertial frame $\mathcal R_i$ , or equivalently, in frame $\mathcal R$: $[\mathbf a]_{\mathcal R_i} = [\mathbf a]_{\mathcal R}$. With the notations of definition \ref{def:dcm}, its projection in $\mathcal R_b$ reads: 
\begin{equation} \label{eq:a_proj}
    [\mathbf a]_{\mathcal R_b} = \mathbf P^T_{b/i}(\boldsymbol \theta^{\mathcal B})[\mathbf a]_{\mathcal R_i} \;.
\end{equation}

\begin{Definition}{Direction Cosine Matrix:}\label{def:dcm} \newline
The Direction Cosine Matrix (DCM) between the body's frame $\mathcal R_b=(O,\mathbf{x}_b,\mathbf{y}_b,\mathbf{z}_b)$ and the frame $\mathcal R_i$, containing the coordinates of vectors $\mathbf{x}_b$, $\mathbf{y}_b$, $\mathbf{z}_b$ expressed in frame $\mathcal R_i$, is noted $\mathbf P_{b/i}(\boldsymbol \theta^{\mathcal B})$.
The inverse function, which converts a DCM $\mathbf{P}_{b/i}$ into the equivalent Euler angles, is noted $\Theta(\mathbf{P}_{b/i}(\boldsymbol \theta^{\mathcal B}))$.
\end{Definition}

\begin{Definition}{From Euler angles rates to angular velocity:}\label{def:pqr}
\newline
The relationship between the body frame angular velocity vector and the rate of change of Euler angles is: 
 \begin{equation}
 [\boldsymbol{\omega}^{\mathcal B}]_{\mathcal R_b} = \boldsymbol \Gamma (\boldsymbol \theta^{\mathcal B}) \dot{\boldsymbol \theta}^{\mathcal B}
 \end{equation}
 where $\boldsymbol \Gamma (\boldsymbol \theta^{\mathcal B})$ depends on the chosen Euler sequence and expresses the relation between the angular velocty vector and the rate of change of Euler angles \cite{zipfel2014modeling}.
\end{Definition}

\section{Connection with a revolute joint}

\label{sec:3_interconnection}

In this section, we consider two bodies $\mathcal A$ and $\mathcal B$ interconnected with a revolute joint (one degree of freedom in rotation).

\subsection{Change of frame}

\label{sec:change_of_frame}

Since the equations describing the motion of $\mathcal B$ (respectively $\mathcal A$) are projected in the reference frame $\mathcal R_b$ (respectively $\mathcal R_a$), the change of frame operation is necessary to write the interconnection of $\mathcal A$ and $\mathcal B$.

\begin{Property}{Change of frame:}\label{prop:change_of_frame} \newline
	Given the Direction Cosine Matrix (DCM) $\mathbf{P}_{a/b}$ between two frames $\mathcal R_a$ and $\mathcal R_b$ according to definition \ref{def:dcm}, let us define $\mathbf{P}_{a/b}^{\times 2} = \diag(\mathbf{P}_{a/b},\mathbf{P}_{a/b})$.
	Then:
	\begin{itemize}
		\item For X a dual velocity, acceleration, or wrench vector: $[\mathbf X]_{\mathcal R_b}=\mathbf P_{a/b}^{\times 2}[\mathbf X]_{\mathcal R_a}$
		\item Direct dynamics model: $\mathbf P_{a/b}^{\times 2}[\mathbf D^{\mathcal A}_P]_{\mathcal R_a}{\mathbf P_{a/b}^{\times 2}}^T=[\mathbf D^{\mathcal A}_P]_{\mathcal R_b}$
		\item $\mathbf P_{b/a} = \mathbf P_{a/b}^{-1} = \mathbf P_{a/b}^{T}$
	\end{itemize}
\end{Property}

\subsection{Model of the revolute joint}

Let $\theta$, $\dot{\theta}$, $\ddot\theta$ be the angular configuration, rate and acceleration inside the revolute joint between bodies $\mathcal{B}$ and $\mathcal{A}$ at the connection point $P$, $\mathbf r$ the vector of unit norm aligned with the joint's axis, and $T_r$ the driving torque along $\mathbf r$.
The revolute joint $\mathcal J$ is modeled as a body with two ports (it is connected to $\mathcal A$ and $\mathcal B$), to which are added an input $\ddot \theta$ and an output $T_r$.
It is assumed that $\mathcal J$ is a mass-less body attached to the body $\mathcal A$, with a matrix of inertia $\mathbf{J}^{\mathcal{J}}_P = J^{\mathcal{J}} \mathbf r \mathbf r^T$. From equation \eqref{eq:Newton_euler_b}, the dynamic model of $\mathcal J$ reads:

\begin{equation} \label{eq:model_R}
\resizebox{\hsize}{!}{$
\begin{aligned}
    \mathbf W_{\mathcal B/\mathcal{J},P}+\mathbf W_{\mathcal{A/J},P}
& =\left[\begin{array}{cc}
\mathbf 0_3 & \mathbf 0_3 \\ \mathbf 0_3 & \mathbf{J}^{\mathcal{J}}_P
\end{array}\right]
\left[\begin{array}{c}
\dot{\mathbf v}^{\mathcal A}_P + \mathbf a\\
 \boldsymbol{\dot{\omega}}^{\mathcal A}
\end{array} \right] \\
& \; \; \; \; + \left[\begin{array}{c}
\mathbf 0_{3\times 1}\\
 (^*\boldsymbol\omega^{\mathcal A})\mathbf{J}^{\mathcal{J}} \boldsymbol\omega^{\mathcal A}
\end{array} \right] \\
& = \left[\begin{array}{c}
\mathbf 0_{3\times 1}\\
\mathbf{J}^{\mathcal{J}}_P \left(\boldsymbol{\dot{\omega}}^{\mathcal B} + \ddot{\theta} \mathbf r \right) + (^*\boldsymbol\omega^{\mathcal A}) \mathbf{J}^{\mathcal{J}}_P \boldsymbol\omega^{\mathcal A}
\end{array} \right].
\end{aligned} $}
\end{equation}

The driving torque $T_r$ is the projection of the torque $\mathbf T_{B/\mathcal J,P}$ applied by $\mathcal B$ on $\mathcal{J}$ at $P$ along $\mathbf r$:
\begin{equation}\label{eq:Cm}
\begin{aligned}
T_r&=\mathbf r_{\mathrm 6}^T\mathbf W_{B/\mathcal{J},P} \\
&= J^{\mathcal J} \left( \mathbf r^T \boldsymbol{\dot{\omega}}^{\mathcal B} + \ddot{\theta} \right) -\mathbf r_6^T\mathbf W_{\mathcal{A/J},P} + \underbrace{\mathbf r^T (^*\boldsymbol\omega^{\mathcal A})\mathbf J^{\mathcal{J}}_P \boldsymbol\omega^{\mathcal A}}_{=0} \;.
\end{aligned}
\end{equation}
where $\mathbf r_{\mathrm 6}= \begin{bsmallmatrix} \mathbf 0_{3\times1} \\ \mathbf r \end{bsmallmatrix}$.
In most applications, it is preferred to invert the channel from $(\theta,\dot\theta,\ddot \theta)$ to $T_r$ to take into account a driving mechanism actuating the revolute joint:
\begin{equation} \label{eq:ddot_theta}
\ddot{\theta} =\frac{1}{J^{\mathcal J}}\left(T_r+\mathbf r_6^T\mathbf W_{\mathcal{A/J},P}  \right)-\mathbf r^T \boldsymbol{\dot{\omega}}^{\mathcal B} \;.
\end{equation}

The motion vector, projected in each body's frame, is transformed through the revolute joint as follows.

\begin{Property}{Transformation of the motion vector through a revolute joint between two bodies:}\label{prop:motion_vector_revolute} \newline
The motion vector at point $P$ can be expressed from body $\mathcal B$ to body $\mathcal A$, connected at point $P$ with a revolute joint:
\begin{equation*}
\resizebox{\hsize}{!}{$
	[\mathbf{m}^{\mathcal A}_P]_{\mathcal R_a } =
	\left[\begin{array}{c}
	[\mathbf a]_{\mathcal R_a} \\\
	[\boldsymbol{\mathbf x''}^{\mathcal A}_{ P}]_{\mathcal R_a}\\\
	[\boldsymbol{\mathbf x'}^{\mathcal A}_{ P}]_{\mathcal R_a} \\\
	[\boldsymbol{\mathbf x}^{\mathcal A}_{P}]_{\mathcal R_a}
	\end{array}\right]
	=
	\left[\begin{array}{c}
	\mathbf P_{b/a}(\theta) [\mathbf a]_{\mathcal R_b} \\\
	\mathbf P_{b/a}^{\times 2}(\theta) [\boldsymbol{\mathbf x''}^{\mathcal B}_{ P}]_{\mathcal R_a} +\ddot{\theta} [\mathbf r_6]_{\mathcal R_a } \\\
	\mathbf P_{b/a}^{\times 2}(\theta)[\boldsymbol{\mathbf x'}^{\mathcal B}_{ P}]_{\mathcal R_a} +\dot{\theta} [\mathbf r_6]_{\mathcal R_a } \\\
	\begin{bsmallmatrix} \mathbf P_{b/a}(\theta)[\overrightarrow{OP}]_{\mathcal R_a} \\ \Theta_{b/a}^{\mathcal J}(\boldsymbol \theta^{\mathcal B}, \theta) \end{bsmallmatrix}
	\end{array}\right] $}
\end{equation*}
where $\boldsymbol \theta^{\mathcal A} = \Theta^{\mathcal J}_{b/a}(\boldsymbol \theta^{\mathcal B}, \theta)$ is defined as $\Theta^{\mathcal J}_{b/a}(\boldsymbol \theta^{\mathcal B}, \theta) = \Theta \left(\mathbf{P}_{b/i}(\boldsymbol \theta^{\mathcal B}) \mathbf{P}_{b/a} (\theta) \right)$.
\end{Property}

\section{Linearization of the individual models}

\label{sec:4_linearization}

In this section, the equations describing the equilibrium and the linear variations around the equilibrium are derived individually for each model of rigid body and revolute joint obtained in Sections \ref{sec:2_rigid_body} and \ref{sec:3_interconnection}. This way, the parametric dependencies are analytically derived on simple models; it is shown why this step is necessary to capture the LFT dependency on certain parameters of interest.

\subsection{Equilibrium}

The system is said to be at equilibrium when it has no motion in the reference frame $\mathcal R$. For a body $\mathcal B$ and a point $P$, it corresponds to: $\left\lbrace \boldsymbol{\mathbf x}^{\mathcal B}_P = \overline{\boldsymbol{\mathbf x}}^{\mathcal B}_P \;, \; \boldsymbol{\mathbf x'}^{\mathcal B}_P = \mathbf 0 \; , \; \boldsymbol{\mathbf x''}^{\mathcal B}_P = \mathbf 0 \right\rbrace$. For a revolute joint $\mathcal J$, it corresponds to $\left\lbrace \theta = \bar\theta \;, \; \dot\theta = 0 \; , \; \ddot\theta  =  0 \right\rbrace$. The Euler angles at equilibrium are noted $\mybar{\boldsymbol \theta}^{\mathcal B}$.
Around the equilibrium, the vectors defined in Section \ref{sec:2A}, projected in $\mathcal R_b$, verify to the first-order: 
\begin{equation} \label{eq:linearized_states}
\left\lbrace\begin{array}{l}
    \delta [\boldsymbol{\mathbf x''}^{\mathcal B}_P]_{\mathcal R_b}=\frac{\mathrm d (\delta [\mathbf x'^{\mathcal B}_P]_{\mathcal R_b})}{\mathrm d t} \\
    \delta [\boldsymbol{\mathbf x'}^{\mathcal B}_P]_{\mathcal R_b}= \diag \left( \mathbf I_3, \boldsymbol \Gamma (\mybar{\boldsymbol\theta}^{\mathcal B}) \right)  \frac{\mathrm d (\delta [\mathbf x^{\mathcal B}_P]_{\mathcal R_b})}{\mathrm d t}
\end{array} \right.
\end{equation}
and the linearized motion vector projected in $\mathcal R_b$ is:
\begin{equation}
    \delta \mathbf{m}^{\mathcal B}_P = \left[ \delta [\mathbf a]_{\mathcal R_b}^T \, , \, \delta [\boldsymbol{\mathbf x''}^{\mathcal B}_P]_{\mathcal R_b}^T \, , \, \delta [\boldsymbol{\mathbf x'}^{\mathcal B}_P]_{\mathcal R_b}^T \, , \, \delta [\boldsymbol{\mathbf x}^{\mathcal B}_P]_{\mathcal R_b}^T \right]^T \;.
\end{equation}

\subsection{Linearized model of the rigid body}

\label{sec:4_linearization_rigid}

Equation \eqref{eq:NE_short_g_projected} is evaluated at equilibrium:
\begin{equation} \label{eq:NE_short_g_equilibrium}
[\overline{\mathbf W}^{\mathcal B}_{P}]_{\mathcal R_b} = 
[\textbf{D}_P^{\mathcal B}]_{\mathcal R_b}  [\overline{\mathbf a}_{\mathrm 6}]_{\mathcal R_b}
\end{equation}
with
\begin{equation}  \label{eq:a_Rb}
    [\overline{\mathbf a}_{\mathrm 6}]_{\mathcal R_b}= \left[ \begin{array}{c} \mathbf P^T_{b/i} (\mybar{\boldsymbol\theta}^{\mathcal B})[\mathbf a]_{R_i} \\ \mathbf 0_{3\times1} \end{array} \right] \;.
\end{equation}
Equation \eqref{eq:NE_short_g_equilibrium} shows that the wrench applied to $\mathcal B$ at equilibrium depends on the DCM $\mathbf P_{b/i} (\mybar{\boldsymbol\theta}^{\mathcal B})$ and on the direct dynamics model $[\textbf{D}_P^{\mathcal B}]_{\mathcal R_b}$ of body $\mathcal B$, which can be an LFT of the following parameters: mass, matrix of inertia, position of the center of gravity $B$ relatively to point $P$. This observation induces that the internal wrenches of the multibody system may be LFTs of these parameters, which will be of importance when linearizing the model of revolute joint in Section \ref{sec:4_linearization_joint}.

Equation \eqref{eq:NE_short_g_projected} is linearized around the equilibrium:
\begin{equation} \label{eq:NE_short_g_variations}
\delta [\mathbf W^{\mathcal B}_{P}]_{\mathcal R_b} = 
[\textbf{D}_P^{\mathcal B}]_{\mathcal R_b}\left( \delta [\boldsymbol{\mathbf x''}^{\mathcal B}_P]_{\mathcal R_b} + \delta [\mathbf a_{\mathrm 6}]_{\mathcal R_b}\right) \;.
\end{equation}
\textit{Remark: including the acceleration vector $\mathbf a$ in the motion vector $\mathbf{m}^{\mathcal B}_P$ allows equation \eqref{eq:NE_short_g_variations} to be linear in the linearized motion vector $\delta [\mathbf{m}^{\mathcal B}_P]_{\mathcal R_b}$, and thus to be compliant with LFT formalism. This is possible because $\delta [\mathbf a]_{\mathcal R_b}$ can be propagated through the revolute joints (see the linearization of property \ref{prop:motion_vector_revolute}, in Section \ref{sec:4_linearization_joint}). If it was not the case, we should instead write $\delta [\mathbf a]_{\mathcal R_b}$ as: 
\begin{equation} \label{eq:remark}
    \delta [\mathbf a]_{\mathcal R_b} = \left. \frac{\mathrm d [\mathbf a]_{\mathcal R_b}}{\mathrm d \boldsymbol\theta^{\mathcal B}}\right|_{\mathrm{eq}} \delta \boldsymbol\theta^{\mathcal B}
\end{equation}
but the matrix $\left. \frac{\mathrm d [\mathbf a]_{\mathcal R_b}}{\mathrm d \boldsymbol\theta^{\mathcal B}}\right|_{\mathrm{eq}}$ cannot generally be obtained as an LFT because it depends on Euler angles (see Appendix).}

The transport of the motion vector (following property \ref{prop:kinematic_transport}) can be linearized around the equilibrium:
\begin{equation} \label{eq:transport_motion_vector_rigid_variations}
\left\lbrace\begin{array}{l}
	\delta [\mathbf{m}^{\mathcal B}_P]_{\mathcal R_b} = [\boldsymbol{\Upsilon}_{PC}]_{\mathcal R_b} \delta [\mathbf{m}^{\mathcal B}_C]_{\mathcal R_b} \\\
	[\boldsymbol{\Upsilon}_{PC}]_{\mathcal R_b} = \diag \left( \mathbf I_3 \, , \, [\boldsymbol{\tau}_{PC}]_{\mathcal R_b}\, , \,[\boldsymbol{\tau}_{PC}]_{\mathcal R_b}\, , \, \mathbf I_6 \right)
\end{array}\right. \;.
\end{equation}

\begin{figure*}[htbp]
\centering
\subfloat[\label{fig:schema_rigide_2}]{\includegraphics[scale=.8]{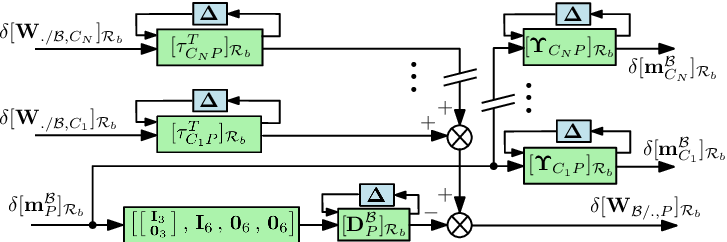}}\hfill
\subfloat[\label{fig:schema_rigide_1}]{\includegraphics[scale=.8]{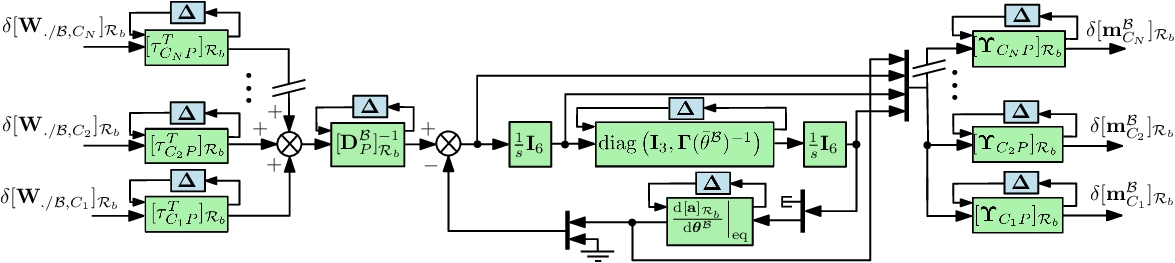}}\hfill
\caption{Linearized model of a rigid body $\mathcal B$: (a) Inverse dynamics, (b) Forward dynamics} \label{fig:linearized_rigid_body}
\end{figure*}

Finally, consider a body $\mathcal B$ where the motion is imposed at parent port $P$, and external wrenches $\mathbf W_{./\mathcal B,C_i}$ are applied at N child ports $C_i$.
From equations \eqref{eq:NE_short_g_variations} and \eqref{eq:transport_motion_vector_rigid_variations}, the linearized inverse dynamics LFT model is represented by the block-diagram in Fig.~\ref{fig:schema_rigide_2}.
Using additionally equation \eqref{eq:linearized_states}, the linearized 12th-order forward dynamics LFT model is represented by the block-diagram in Fig.~\ref{fig:schema_rigide_1} for a body with N ports $C_i$ where only external wrenches are applied (no imposed motion). The green and blue blocks represent the nominal models and the $\bm \Delta$ operators respectively.

\textit{Remark: a multibody system has a base which is either the ground (imposed motion) or a body with forward dynamics (no imposed motion, the equilibrium is determined by the wrenches). In the latter case, the orientation of the base at equilibrium is explicitely defined. Then, the matrix $\left. \frac{\mathrm d [\mathbf a]_{\mathcal R_b}}{\mathrm d \boldsymbol\theta^{\mathcal B}}\right|_{\mathrm{eq}}$ can be obtained as an LFT of the Euler angles. However, for the inverse dynamics, the Euler angles are propagated from the base to the body through the joints, and since this transformation is not compliant with the LFT (see the discussion in Appendix), it is necessary to use the acceleration propagated with the motion vector.}

\subsection{Linearized model of the revolute joint}

\label{sec:4_linearization_joint}

The linearization of the revolute joint must take into account the dependency of the DCM $\mathbf P_{a/b}$ on the variable $\theta$: if $\mathbf X$ is a vector such that $[\mathbf X]_{\mathcal R_a} =\mathbf P_{b/a} (\theta) [\mathbf X]_{\mathcal R_b}$:
\begin{equation} \label{eq:linearized_change_of_frame}
    \delta [\mathbf X]_{\mathcal R_a} =\mathbf P_{b/a} (\bar \theta) \delta [\mathbf X]_{\mathcal R_b} + \underbrace{\left.\frac{\mathrm d \mathbf P_{b/a}}{\mathrm d \theta}\right|_{\mathrm{eq}}}_{=(\mathbf r^*)\mathbf P_{b/a}(\bar\theta)} \delta \theta [\overline{\mathbf X}]_{\mathcal R_b} \;.
\end{equation}
\textit{Remark: the DCM $\mathbf P_{b/a} (\bar \theta)$ can be expressed as an LFT of the parameter $t=\tan(\bar\theta/2)$ (respectively $t=\tan(\bar\theta/4)$) with 2 (respectively 4) occurrences (cf. \cite[p. 191 to 195]{Dubanchet2016}}).

The projections of equation \eqref{eq:model_R} in the frame $\mathcal R_b$ and of equation \eqref{eq:ddot_theta} along $\mathbf r$ read:
\begin{equation} \label{eq:model_R_proj}
\resizebox{\hsize}{!}{$
\left\lbrace \begin{array}{l}
[\mathbf W_{\mathcal B/\mathcal J,P}]_{\mathcal R_b} + \mathbf P^{\times 2}_{a/b} (\theta) [\mathbf W_{\mathcal A/\mathcal J,P}]_{\mathcal R_a} \\\
     \quad \quad = \left[ \begin{array}{c}
    \mathbf 0_{3\times1}\\\
    [\mathbf{J}^{\mathcal{J}}_P]_{\mathcal R_b} \left( [\boldsymbol{\dot{\omega}}^{\mathcal B}]_{\mathcal R_b} +\ddot{\theta}[\mathbf r]_{\mathcal R_b} \right) +[ (^*\boldsymbol\omega^{\mathcal A}) \mathbf{J}^{\mathcal{J}}_P \boldsymbol\omega^{\mathcal A}]_{\mathcal R_b}
    \end{array}\right] \\\
    \ddot{\theta} = \frac{1}{J^{\mathcal J}}\left(T_r+[\mathbf r_6^T]_{\mathcal R_a} [\mathbf W_{\mathcal{A/J},P}]_{\mathcal R_a}  \right) 
     - [\mathbf r^T]_{\mathcal R_b} [\boldsymbol{\dot{\omega}}^{\mathcal B}]_{\mathcal R_b}
\end{array} \right. . $}
\end{equation}
It can be noted that $[\mathbf r]_{\mathcal R_b}$ and $[\mathbf{J}^{\mathcal{J}}_P]_{\mathcal R_b}$ are independent from $\theta$. Indeed, noting $\mathbf R (\theta)$ the rotation matrix around $\mathbf r$, the DCM reads $\mathbf P_{b/a} (\theta) = \mathbf R (\theta) \mathbf P_{b/a} (0)$. Then:
\begin{equation}
\begin{aligned}
    [\mathbf r]_{\mathcal R_b} = \mathbf P^T_{b/a} (0) \underbrace{\mathbf R^T (\theta)  [\mathbf r]_{\mathcal R_a}}_{=[\mathbf r]_{\mathcal R_a}}
\end{aligned}
\end{equation}
and
\begin{equation}
\begin{aligned}
    [\mathbf{J}^{\mathcal{J}}_P]_{\mathcal R_b} &= \mathbf P_{b/a}^T (\theta) [\mathbf{J}^{\mathcal{J}}_P]_{\mathcal R_a} \mathbf P_{b/a} (\theta) \\
    &= J^{\mathcal{J}} \mathbf P_{b/a}^T (0) \underbrace{\mathbf R (\theta)^T [\mathbf r]_{\mathcal R_a}}_{=[\mathbf r]_{\mathcal R_a}} \underbrace{[\mathbf r^T]_{\mathcal R_a} \mathbf R (\theta)}_{=[\mathbf r^T]_{\mathcal R_a}} \mathbf P_{b/a} (0) \\
    &= \mathbf P_{b/a}^T(0) [\mathbf{J}^{\mathcal{J}}_P]_{\mathcal R_a} \mathbf P_{b/a} (0) \; .
\end{aligned}
\end{equation}

Equation \eqref{eq:model_R_proj} is evaluated at equilibrium:
\begin{equation} \label{eq:model_R_equilibrium}
\left\lbrace \begin{array}{l}
    [\overline{\mathbf W}_{\mathcal J/\mathcal B,P}]_{\mathcal R_b} = \mathbf P^{\times 2}_{a/b}(\bar \theta) [\overline{\mathbf W}_{\mathcal{A/J},P}]_{\mathcal R_a} \\\
    0 = \mybar T_r+[\mathbf r_6^T]_{\mathcal R_a} [\overline{\mathbf W}_{\mathcal{A/J},P}]_{\mathcal R_a}
\end{array} \right.
\end{equation}
and linearized around the equilibrium:
\begin{equation} \label{eq:modele_revolute_joint}
\left\lbrace \begin{array}{l}
     \delta [\mathbf W_{\mathcal B/\mathcal J,P}]_{\mathcal R_b} + \begin{bsmallmatrix}(\mathbf r^*) & \mathbf 0\\ \mathbf 0 & (\mathbf r^*) \end{bsmallmatrix} \mathbf P^{\times2}_{b/a}(\bar\theta) [\overline{\mathbf W}_{\mathcal{A/J},P}]_{\mathcal R_a} \delta \theta  \\\
     \quad \quad + \mathbf P^{\times 2}_{a/b} (\bar \theta) \delta [\mathbf W_{\mathcal A/\mathcal J,P}]_{\mathcal R_a} \\\ \quad \quad = 
     \left[ \begin{array}{c}
        \mathbf 0_{3\times1}\\\
        [\mathbf{J}^{\mathcal{J}}_P]_{\mathcal R_b} \left(\delta [\boldsymbol{\dot{\omega}}^{\mathcal B}]_{\mathcal R_b} + \delta \ddot{\theta}[\mathbf r]_{\mathcal R_b} \right)
    \end{array}\right] \\
    \delta \ddot{\theta}  = \frac{1}{J^{\mathcal J}}\left( \delta  T_r+ [\mathbf r_6^T]_{\mathcal R_a} \delta [\mathbf W_{\mathcal{A/J},P}]_{\mathcal R_a}  \right) \\
    \qquad \qquad - [\mathbf r^T]_{\mathcal R_b} \delta [\boldsymbol{\dot{\omega}}^{\mathcal B}]_{\mathcal R_b} 
\end{array} \right. \;.
\end{equation}
Equation \eqref{eq:modele_revolute_joint} shows that the wrench applied to the joint at equilibrium, represented by the vector $[\overline{\mathbf W}_{\mathcal{A/J},P}]_{\mathcal R_a}$, introduces a stiffness in the motion of the revolute joint (factor multiplying $\delta \theta$). In addition to some possible wrenches applied to the system and defining the equilibrium (such as a buoyant force compensating for the gravity acceleration in the case of a stratospheric balloon, or a thrust providing the acceleration in the case of a launcher), the wrench $[\overline{\mathbf W}_{\mathcal{A/J},P}]_{\mathcal R_a}$ results from the wrenches applied by rigid bodies given by equation \eqref{eq:NE_short_g_equilibrium}, and it must be evaluated at equilibrium before the linearization.
Therefore, following the discussion on equation~\eqref{eq:NE_short_g_equilibrium}, $[\overline{\mathbf W}_{\mathcal{A/J},P}]_{\mathcal R_a}$ may depend on some parameters of interest (masses, lengths, etc). A numerical evaluation of the trim point, as it is done with current available software, is not adequate to capture it as an LFT (it will only evaluate a single, nominal configuration); on the contrary, evaluating $[\overline{\mathbf W}_{\mathcal{A/J},P}]_{\mathcal R_a}$ while preserving its LFT structure allows to correctly re-inject it in the linearized model of the revolute joint. This observation justifies the need for the analytical derivation of the trim conditions and analytical linearization presented in this section, as well as for the dedicated procedure presented in Section \ref{sec:5_procedure}.

\textit{Example 1: Pendulum -- Consider a pendulum around its stable equilibrium, composed of a revolute joint, a mass-less link, and an point mass, and assume that the mass is uncertain and represented by an LFT. The stiffness is proportional to the mass, and must be computed as an LFT to be re-injected in the linearized model of the revolute joint. It is worth emphasizing that a system as simple as the pendulum has a parameter-dependent equilibrium in the sense of this paper, even though the equilibrium angle is fixed, and must be treated with the proposed approach to derive a multibody LFT model.}

\textit{Example 2: Robotic arm -- Consider a robotic arm with several bodies and revolute joints. It is sought to derive a LPV model where the scheduling parameters, whose variations are isolated with the LFT formalism, are the equilibrium angle of each joint. In addition to the rigid bodies parameters (masses, etc), the internal wrenches also depend on the equilibrium angles of the bodies. Therefore, it is first necessary to derive the DCM $\mathbf P_{b/i} (\mybar{\boldsymbol\theta}^{\mathcal B})$ of each body as the product of the individual DCMs of the revolute joints, which are LFT-LPV rotation matrices. Then, the wrenches are evaluated as LFTs of the equilibrium angles and rigid bodies parameters, and finally re-injected in the linearized models of the revolute joints.}

The transformation of the motion vector (property \ref{prop:motion_vector_revolute}) is also linearized:
\begin{equation} \label{eq:motion_vector_revolute_joint}
\left\lbrace
\begin{array}{l}
    \delta [\mathbf a]_{\mathcal R_a} = (\mathbf r^*) \mathbf P_{b/a} (\bar \theta) [\mybar{\mathbf a}]_{\mathcal R_a} \delta \theta + \mathbf P_{b/a} (\bar \theta) \delta [\mathbf a]_{\mathcal R_b}\\
    \delta [\boldsymbol{\mathbf x''}^{\mathcal A}_P]_{\mathcal R_a} = \mathbf P^{\times2}_{b/a} (\bar \theta) \delta [\boldsymbol{\mathbf x''}^{\mathcal B}_P]_{\mathcal R_b} + \delta \ddot \theta [\mathbf r_6]_{\mathcal R_a} \\
    \delta [\boldsymbol{\mathbf x'}^{\mathcal A}_P]_{\mathcal R_a} = \mathbf P^{\times2}_{b/a} (\bar \theta) [\boldsymbol{\mathbf x'}^{\mathcal B}_P]_{\mathcal R_b} + \delta \dot \theta [\mathbf r_6]_{\mathcal R_a} \\
    \delta [\boldsymbol{\mathbf x}^{\mathcal A}_P ]_{\mathcal R_a} = \left( \begin{array}{cc}
       \mathbf P_{b/a} (\bar \theta)  & \mathbf 0 \\
        \mathbf 0 & \left. \frac{\partial \Theta^{\mathcal J}_{a/b}}{\partial \boldsymbol \theta^{\mathcal B}}\right|_{\mathrm{eq}}
    \end{array} \right) \delta [\boldsymbol{\mathbf x}^{\mathcal B}_P ]_{\mathcal R_b} \\
    \qquad \qquad + 
    \left( \begin{array}{c} (\mathbf r^*)\mathbf P_{b/a}(\bar\theta) [\mybar{OP}]_{\mathcal R_b} \\ \left. \frac{\partial \Theta^{\mathcal J}_{a/b}}{\partial \theta}\right|_{\mathrm{eq}} \end{array} \right) \delta \theta
\end{array} \right.
\end{equation}
Once again, the position vector at equilibrium $[\mybar{OP}]_{\mathcal R_b}$ must be computed analytically because it may have LFT dependency on the lengths or angles at equilibrium.

From equations \eqref{eq:modele_revolute_joint} and \eqref{eq:motion_vector_revolute_joint}, the linearized model of the revolute joint takes $\delta [\mathbf W_{\mathcal A/\mathcal J,P}]_{\mathcal R_a}$, $\delta [\mathbf m^{\mathcal B}_P]_{\mathcal R_b}$ and $\delta T_r$ as inputs, and returns $\delta [\mathbf m^{\mathcal A}_P]_{\mathcal R_a}$, $\delta [\mathbf W_{\mathcal J/\mathcal B,P}]_{\mathcal R_b}$ and $(\delta \ddot \theta, \delta \dot \theta, \delta \theta)$ as outputs. With the proposed evaluation of the vectors $[\overline{\mathbf W}_{\mathcal{A/J},P}]_{\mathcal R_a}$ and $[\mybar{OP}]_{\mathcal R_b}$ as LFT models, the linearized revolute joint is also an LFT model except for the gains $\left. \frac{\partial \Theta^{\mathcal J}_{a/b}}{\partial \theta}\right|_{\mathrm{eq}}$ and $\left. \frac{\partial \Theta^{\mathcal J}_{a/b}}{\partial \boldsymbol \theta^{\mathcal B}}\right|_{\mathrm{eq}}$ which are used to propagate Euler angles (see the discussion in Appendix).



\section{Assembly, trim and linearization algorithm}

\begin{figure}[!ht]
	\centering
	\includegraphics[width=.9\columnwidth]{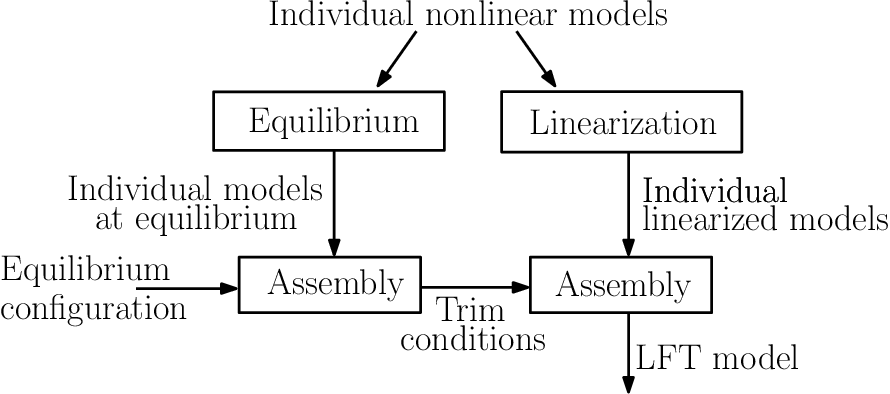}
	\caption{Assembly, trim and linearization algorithm}
	 \label{fig:assembly}
\end{figure}

As discussed in Section \ref{sec:4_linearization}, it is necessary to perform an analytical trimming to preserve the LFT dependencies. This is possible by assembling the model of the structure at equilibrium from the individual models \eqref{eq:NE_short_g_equilibrium} and \eqref{eq:model_R_equilibrium}. Then, the trim conditions, expressed as LFTs, are re-injected in the assembly of the individual linearized models (Fig.\ref{fig:linearized_rigid_body}, equations \eqref{eq:modele_revolute_joint} and \eqref{eq:motion_vector_revolute_joint}). This procedure is schematized in Fig.\ref{fig:assembly}.

\label{sec:5_procedure}

More precisely, let us consider a tree-like structure composed of (i) a base, which is either a parent body described by its forward dynamics ($6$ DOF) or the ground (no DOF), (ii) children bodies described by their inverse dynamics (no additional DOF), and (iii) $n$ revolute joints ($n$ DOF). Each body may be connected to any number of other bodies or joints, as long as there is no closed kinematic loop.

\textbf{Step 1 (Geometry at equilibrium, or forward recurrence):}
This step aims at computing the geometrical trim conditions as LFTs of the parameters of interest: the DCM $\mathbf P_{b/i} (\mybar{\boldsymbol\theta}^{\mathcal B})$ for each body, and the position vector $[\mybar{OP}]_{\mathcal R_b}$ at each revolute joint. These quantities are initially defined at the base (either a parent body or ground) and are propagated from the base to the other bodies and joints. The DCM is transformed at each revolute joint: $\mathbf P_{b/i} (\mybar{\boldsymbol\theta}^{\mathcal B})=\mathbf P_{a/i} (\mybar{\boldsymbol\theta}^{\mathcal A})\mathbf P_{b/a} (\bar\theta)$; and the position vector is transformed at each revolute joint: $[\mybar{OP}]_{\mathcal R_b} = \mathbf P_{a/b} (\bar\theta) [\mybar{OP}]_{\mathcal R_a}$, and at each rigid body: from a port $P$ to a port $C$: $[\mybar{OC}]_{\mathcal R_b} = [\mybar{OP}]_{\mathcal R_b} + [CP]_{\mathcal R_b}$. The DCMs $\mathbf P_{b/a} (\bar\theta)$ and the positions $[CP]_{\mathcal R_b}$ can be LFT models, and these operations preserve the LFT form, hence all $\mathbf P_{b/i} (\mybar{\boldsymbol\theta}^{\mathcal B})$ and $[\mybar{OP}]_{\mathcal R_b}$ are finally obtained as LFT models.

\textbf{Step 2 (Wrenches at equilibrium, or backward recurrence):} 
This step aims at computing the wrenches at equilibrium $[\overline{\mathbf W}_{\mathcal{A/J},P}]_{\mathcal R_a}$ in the revolute joints as LFTs of the parameters of interest. For this, the wrenches are propagated from the outer bodies (end of the open kinematic chain) to the base using the models \eqref{eq:NE_short_g_equilibrium} and \eqref{eq:model_R_equilibrium}, which are also compliant with the LFT formalism (and where the matrices $[\textbf{D}_P^{\mathcal B}]_{\mathcal R_b}$ can be LFT models as well). Note that step 2 requires the DCMs $\mathbf P_{b/i} (\mybar{\boldsymbol\theta}^{\mathcal B})$ computed at step 1 (see equation \eqref{eq:a_Rb}).

\textbf{Step 3 (Linearized model):} Finally, the individual linearized models (Fig.\ref{fig:linearized_rigid_body}, equations \eqref{eq:modele_revolute_joint} and \eqref{eq:motion_vector_revolute_joint}) are assembled while re-injecting the trim conditions obtained as LFT models in steps 1 and 2. Therefore, the resulting model is a fully parameterized LFT model accounting for the parameter-dependent equilibrium.

Since the physical origin of all parameters has been preserved during the whole procedure, the LFT model exactly covers all plants without introducing conservatism or fitting error. In practice, the procedure can be implemented on Matlab-Simulink; in this case, the trim conditions computed as LFT models in steps 1 and 2 are evaluated as input/output transfers after implementation of models \eqref{eq:NE_short_g_equilibrium} and \eqref{eq:model_R_equilibrium} as static LFT models. Since only basic block-diagram manipulations are applied, the procedure can be executed in reasonable time even for complex systems.
However, although the trim conditions calculated in steps 1 and 2 can be expressed with minimal parametric dependency on the parameters of interest, since they are in turn re-injected at step 3, there can be redundant occurrences in the linearized LFT model; reduction techniques can be used to reduce the order of the block $\bm \Delta$ \cite{Varga1999}.


\section{Application example}

\label{sec:6_application}

\subsection{Presentation of the system}

\begin{figure}[!ht]
	\centering
	\includegraphics[width=.7\columnwidth]{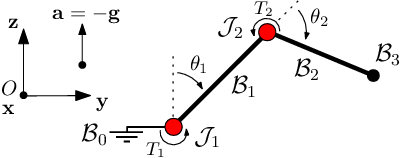}
	\caption{Two-link robotic robotic arm}
	 \label{fig:robotic_arm}
\end{figure}

The two-link robotic arm presented in Fig.~\ref{fig:robotic_arm} is subject to the gravity represented by the vector $\mathbf g$, which is equivalent to an acceleration $\mathbf a = - \mathbf g$ in the proposed approach.
The reference frame in acceleration is noted $\mathcal R = (O, \mathbf x, \mathbf y, \mathbf z)$.
The arm is composed of 3 bodies $\mathcal B_1$, $\mathcal B_2$, and $\mathcal B_3$.
The revolute joints $\mathcal J_1$ and $\mathcal J_2$, which allow the rotation around $\mathbf x$, are actuated with torques $T_1$ and $T_2$. $\mathcal B_3$ is a point mass representing the end-effector carrying a load, and is rigidly connected to $\mathcal B_2$ (no degree of freedom).
The characteristics of the rigid bodies are indicated in Table~\ref{tab:numbers}.
The position of the center of gravity (CoG) is the distance of the CoG from the body's left tip (in Fig.~\ref{fig:robotic_arm}), normalized by the length of the body.
Uncertainties of $\pm 20 \%$ have been set on some parameters.
The scheduling parameters $t_1=\tan(\bar\theta_1/2)$ and $t_2=\tan(\bar\theta_2/2)$ are defined as uncertain parameters in the revolute joints blocks. 

\begin{table}[hbt!]
\caption{\label{tab:numbers} Physical parameters of the robotic arm}
\centering
\resizebox{\columnwidth}{!}{
\begin{tabular}{c|ccc}
& $\mathcal B_1$ & $\mathcal B_2$ & $\mathcal B_3$   \\ \hline \hline
Mass $m_i$ (kg) &  3 ($\pm$ 20\%) & 2 & 5 ($\pm$ 20\%) \\
Moment of inertia $J_i$ (kg.m\textsuperscript{2}) &  0.2 ($\pm$ 20\%) & 0.1 & 0  \\
Length $L_i$ (m) &  1 & 1 ($\pm$ 20\%) &  0  \\
Position of the CoG $\rho_i$ (-) &  0.3 ($\pm$ 20\%) & 0.5 & 0\\
\end{tabular}
}
\end{table}

\subsection{Multibody LFT modeling}

\begin{figure}[!ht]
	\centering
	\includegraphics[width=0.9\columnwidth]{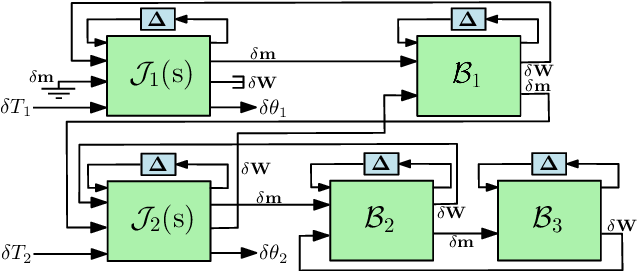}
	\caption{Multibody LFT model of the robotic arm}
	 \label{fig:block_diagram}
\end{figure}

The proposed approach is implemented on \textsc{Matlab} with the \textit{robust control toolbox}. The uncertain and scheduling parameters are declared with the routine \verb!ureal!.
The trim conditions (DCMs, position vectors, wrenches) are evaluated with the routine \verb!ulinearize! as static input/output transfers in separated \textsc{Simulink} files, where the individual static LFT models of each body at equilibrium are assembled (steps 1 and 2).
Once the trim conditions are obtained as LFT models, they are re-injected in the linearized models which are assembled as in Fig.~\ref{fig:block_diagram} (step 3). For readability, it is indicated whether the connections represent a motion vector $\delta \mathbf m$ or a wrench $\delta \mathbf W$, but the full nomenclature adopted in previous sections is omitted. The LFT dependencies of the trim conditions are carried by the blocks $\bm\Delta$ of the revolute joints.
A damping \mbox{$K_d$ = \SI{0.1}{\newton\second\per\radian}} and a stiffness \mbox{$K_p$ = \SI{0.1}{\newton\per\radian}}  are added to the linear models of the revolute joints.
The procedure took 20 seconds on a Intel Core i7 processor.

\vspace{-0.4cm}

\subsection{Comparison with \textit{Simscape Multibody}}

\begin{figure}[!ht]
	\centering
	\includegraphics[width=0.8\columnwidth]{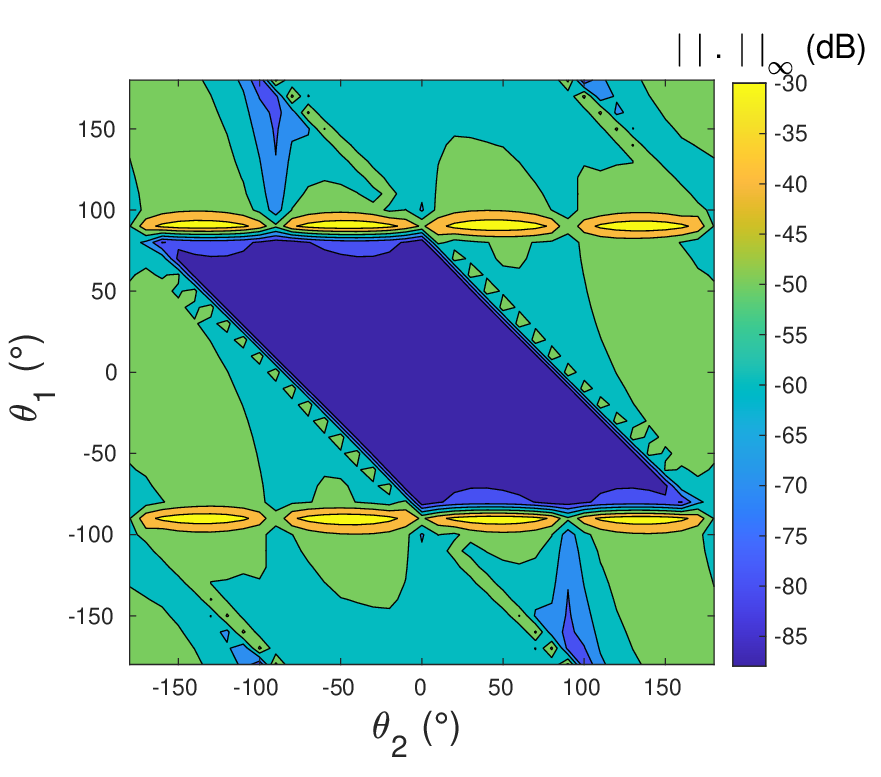}
	\caption{$\mid\mid (\mathbf G_1 - \mathbf  G_2) \mathbf G_2^{-1} \mid\mid_{\infty}$ across angular configurations, where $\mathbf G_1$ is the nominal LFT model (no parametric uncertainty) and $\mathbf G_2$ is the \textit{Simscape Multibody}'s model.}
	 \label{fig:comparison1}
\end{figure}

\begin{figure}[!ht]
	\centering
	\includegraphics[width=\columnwidth]{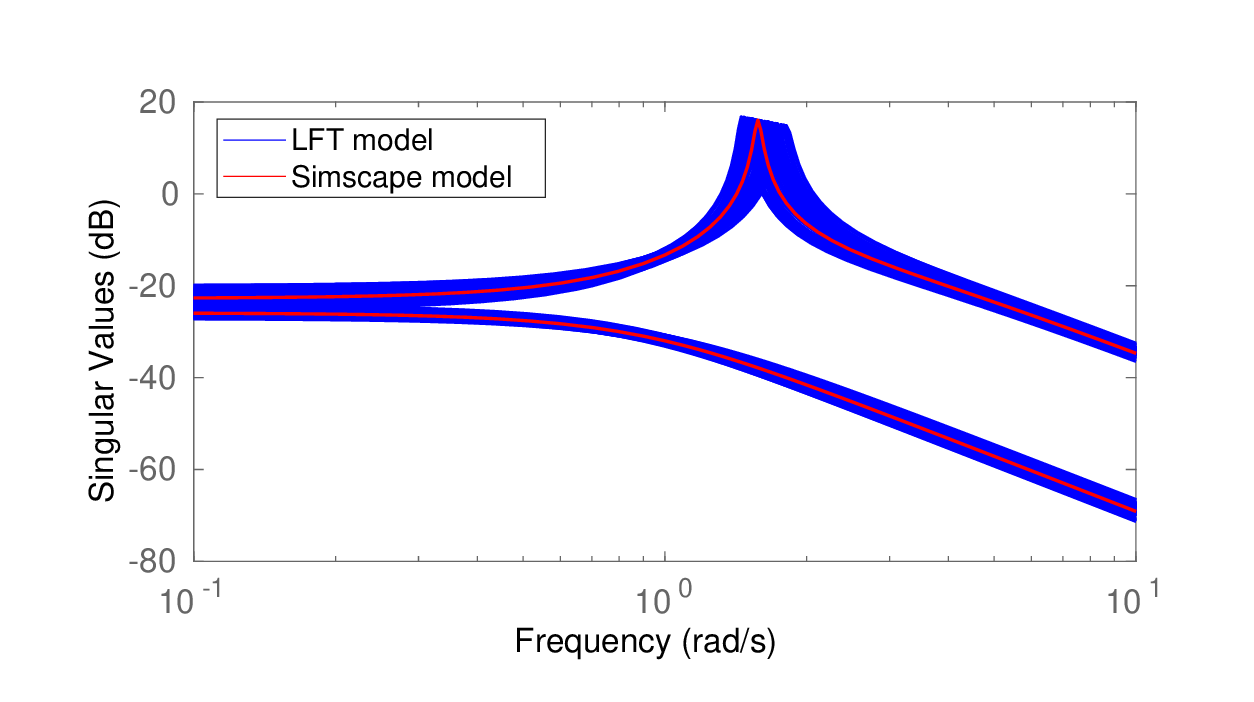}
	\caption{Singular values of $[\delta T_1 \,,\, \delta T_2 ]^T \rightarrow [\delta \theta_1 \,,\, \delta \theta_2]^T$ for $\bar\theta_1=\SI{70}{\degree}$ and $\bar \theta_2=\SI{30}{\degree}$ (300 samples of the LFT model)}
	 \label{fig:comparison2}
\end{figure}

To validate the proposed approach, the model of the same robotic arm is built with \textit{Simscape Multibody} and linearized around the equilibrium. The proposed LFT model matches \textit{Simscape}'s model in the nominal configuration of the uncertain parameters and across all angular configurations, as shown in Fig.~\ref{fig:comparison1}, where the relative error between the two models stays small even in the worst-case configurations around $\bar\theta_1 = \pm \SI{90}{\degree}$. Non nominal configurations were also tested and matched the corresponding \textit{Simscape}'s model.
Moreover, Fig.~\ref{fig:comparison2} presents the singular values of the transfer $[\delta T_1 \,,\, \delta T_2 ]^T \rightarrow [\delta \theta_1 \,,\, \delta \theta_2]^T$ for both models in one angular configuration. Let us emphasize that the proposed LFT model contains all configurations of the scheduling parameters $t_1$ and $t_2$ as well as the parametric uncertainties in one single model, while the \textit{Simscape} model needs to be reevaluated, trimmed and linearized for every geometric or parametric configuration.

\vspace{-0.3cm}

\vspace{-1cm}

\vspace{-0.7cm}

\subsection{Robust LPV control}

To conclude, a robust LPV controller is proposed to illustrate the compatibility of the proposed approach with classical robust control tools and to show the advantages of the LFT model.
The angles are limited to the following operating ranges: $\theta_1 \in \left[  \SI{45}{\degree}, \SI{90}{\degree}  \right]$ and $\theta_2 \in \left[  \SI{45}{\degree}  ,  \SI{135}{\degree} \right]$, and the set of scheduling parameters is noted $\bm\tau = \left\lbrace t_1 , t_2 \right\rbrace$. 

Noting $\delta \bm \theta_{\mathrm{ref}} = [ \delta \theta_1^{\mathrm{ref}} \,,\,
\delta \theta_2^{\mathrm{ref}} ]^T$ the vector of reference angles, $\delta \mathbf e = \delta \bm \theta_{\mathrm{ref}} - [ \delta \theta_1 \,,\, \delta \theta_2 ]^T$, and $\delta \mathbf T =
[ \delta T_1 \,,\, \delta T_2]^T$, the LPV controller $\mathbf K(\mathrm s, \bm\tau)$ is such that: 
\begin{equation}
\delta \mathbf T = \mathbf K(\mathrm s, \bm\tau) \delta \mathbf e \;.
\end{equation}
Let the real matrices of appropriate dimensions $\mathbf A(\bm\tau)$, $\mathbf B(\bm\tau)$, $\mathbf C(\bm\tau)$, $\mathbf D(\bm\tau)$ define the state-space representation of $\mathbf K(\mathrm s, \bm\tau)$.
The scheduling surface $\mathbf S(\bm\tau)$ is defined as:
\begin{equation} 
     \mathbf S(\bm\tau) = \left[ \begin{array}{c|c} \mathbf A(\bm\tau) & \mathbf B(\bm\tau) \\\hline \mathbf C(\bm\tau) & \mathbf D(\bm\tau) \end{array} \right] 
     = \mathbf M_0 + \mathbf M_1  t_1 +  \mathbf M_2 t_2
\end{equation}
where the matrices $\mathbf M_0$, $\mathbf M_1$, $\mathbf M_2$ are to be tuned, and the LPV controller $\mathbf K(\mathrm s, \bm\tau)$ reads:
\begin{equation}
\begin{aligned}
    \mathbf K(\mathrm s, \bm\tau)
    = \mathcal F_u \left( \mathbf S(\bm\tau),  \frac{\mathbf I_{n_s}}{\mathrm s} \right)  
    = \mathcal F_u \left( \mathbf K(\mathrm s), \bm\Delta^K_{\bm\tau}  \right) 
\end{aligned}
\end{equation}
where $\mathcal F_u$ refers to the upper LFT, $n_s$ is the number of states of the controller, and the block $\Delta^K_{\bm\tau}$ isolates the occurrences of $t_1$ and $t_2$.

\begin{figure}[!ht]
	\centering
	\includegraphics[width=\columnwidth]{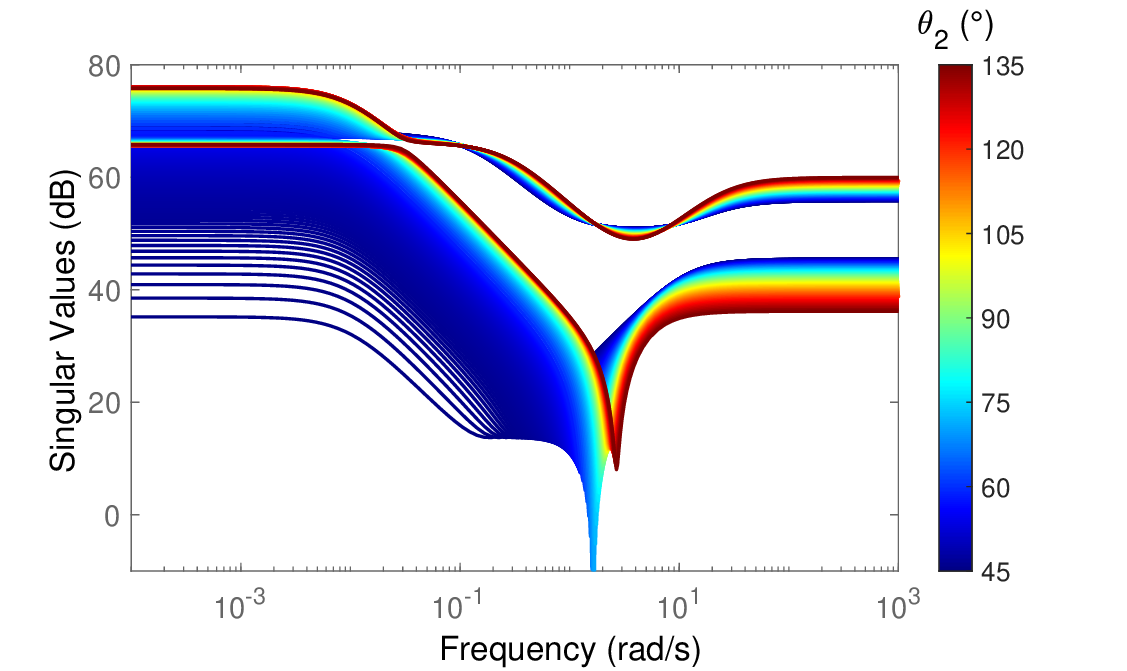}
	\caption{Singular values of $\mathbf K(\mathrm s, \bm\tau)$, $\bar \theta_1=\SI{45}{\degree}$, $\bar \theta_2 \in \left[  \SI{45}{\degree}  ,  \SI{135}{\degree} \right]$}
	 \label{fig:controller}
\end{figure}

The value $n_s=3$ was chosen, and after defining the weighting functions $\mathbf W_{T} = 1/1500 \, \mathbf I_2$ (to limit the actuator's efforts) and $\mathbf W_e (\mathrm s) = \frac{\mathrm s + 1}{2 \mathrm s + 0.02} \, \mathbf I_2$ (to penalize low-frequency tracking error), the robust, structured $\mathcal H_{\infty}$ problem:
\begin{equation}
    \begin{array}{l}
    \underset{\mathbf M_0, \mathbf M_1, \mathbf M_2}{\text{minimize}} \; \; \; \gamma_2 \; \; \; s.t. \; \; 
        \underset{\bm \tau, \bm \Delta}{\max} \left\lbrace \mid \mid \delta \bm \theta_{\mathrm{ref}} \rightarrow \mathbf W_{T} \delta \mathbf T \mid \mid_{\infty} \right\rbrace < \gamma_2\\
        \text{subject to: } \underset{\bm \tau, \bm \Delta}{\max} \left\lbrace \mid \mid \delta \bm \theta_{\mathrm{ref}} \rightarrow \mathbf W_e \delta \mathbf e \mid \mid_{\infty} \right\rbrace < \gamma_1 < 1
    \end{array}
\end{equation}
was solved with \textsc{Matlab} routine \verb!systune!, based on the algorithm presented in \cite{Apkarian2006}. 
A performance ($\gamma_1=0.97$, $\gamma_2=0.69$) was obtained (corresponding to the worst-case $\mathcal H_{\infty}$ norms of the transfers), and Fig.\ref{fig:controller} represents the LPV controller.
Since the proposed modeling approach provided all parametric configurations of both the uncertain and scheduling parameters in one single LFT model, the robustness and the LPV controller synthesis were addressed together in one single control design iteration, and the resulting performance is guaranteed across all parametric configurations.

\section{Conclusion}

After introducing a multibody modeling framework based on Newton-Euler equations, it was shown why a numerical trim computation is not adequate to derive an LFT model, and a specific assembly procedure, based on the linearization of the equations of motion at the substructure level, was proposed to solve this issue. An application to a robotic arm was outlined to show how the proposed approach can be implemented on \textsc{Matlab} and used for control design.

\appendices

\section*{Appendix}

The transformation $\Theta(.)$ from definition \ref{def:dcm} cannot be expressed as an LFT of uncertain or varying Euler angles, because it includes trigonometric functions. Therefore, propagating Euler angles from one body to another (with the function $\Theta^{\mathcal J}_{b/a}$ from property \ref{prop:motion_vector_revolute}) cannot be done while preserving the LFT form.
As a consequence, if Euler angles are defined as output measurements, the corresponding output gains cannot always be obtained as exact LFTs, and rational approximations of $\Theta^{\mathcal J}_{a/b}$ and its derivatives may be necessary (it can be noted that, for problems in a single plane, the transformation $\Theta^{\mathcal J}_{a/b}$ becomes trivial and this issue disappears).
Nonetheless, the dynamical model can always be obtained because the inclusion of the acceleration vector $\mathbf a$ in the motion vector allows to dispense with Euler angles in the equations of the dynamics (see equation \eqref{eq:NE_short_g_variations} in Section \ref{sec:4_linearization_rigid}).

\bibliographystyle{IEEEtran}
\bibliography{library}

\end{document}